\documentclass[twocolumn,preprintnumbers,amssymb,amsmath,aps,floatfix,prc,nofootinbib,superscriptaddress,showpacs]{revtex4-1}

\usepackage{graphicx}
\usepackage{amsmath,amsthm,amssymb}

\usepackage{color}
\usepackage{bm}
\usepackage{lipsum}
\usepackage[bookmarks,
                bookmarksopen = true,
                bookmarksnumbered = true,
                linktocpage,
                colorlinks = true,
                linkcolor = blue,
                urlcolor  = blue,
                citecolor = blue,
                anchorcolor = green,
                hyperindex = true,
                hyperfigures]
                {hyperref}

\begin{document}

\title{Probing medium-induced jet splitting and energy loss in heavy-ion collisions}
\author{Ning-Bo Chang}
\affiliation{Institute of Theoretical Physics, Xinyang Normal University, Xinyang, Henan 464000, China}
\affiliation{Institute of Particle Physics and Key Laboratory of Quark and Lepton Physics (MOE), Central China Normal University, Wuhan, 430079, China}

\author{Shanshan Cao}
\affiliation{Department of Physics and Astronomy, Wayne State University, Detroit, MI 48201, USA}

\author{Guang-You Qin}
\affiliation{Institute of Particle Physics and Key Laboratory of Quark and Lepton Physics (MOE), Central China Normal University, Wuhan, 430079, China}

\date{\today}

\begin{abstract}

The nuclear modification of jet splitting in relativistic heavy-ion collisions at RHIC and the LHC energies is studied based on the higher twist formalism.
Assuming coherent energy loss for the two splitted subjets, a non-monotonic jet energy dependence is found for the nuclear modification of jet splitting function: strongest modification at intermediate jet energies whereas weaker modification for larger or smaller jet energies.
Combined with the smaller size and lower density of the QGP medium at RHIC than at the LHC, this explains the CMS-STAR groomed jet puzzle -- strong nuclear modification of the momentum sharing $z_g$ distribution at the LHC whereas no obvious modification of the $z_g$ distribution at RHIC.
In contrast, the observed nuclear modification pattern of the groomed jet $z_g$ distribution cannot be explained solely by independent energy loss of the two subjets.
Our result may be tested in future measurements of groomed jets with lower jet energies at the LHC and larger jet energies at RHIC, for different angular separations between the two subjets.

\end{abstract}

\maketitle

\section{Introduction}
\label{sec:introduction}

Jet quenching has been regarded as one of the most important tools to study the novel properties of the quark-gluon plasma (QGP) created in relativistic heavy-ion collisions at the Relativistic Heavy-Ion Collider (RHIC) and the Large Hadron Collider (LHC) \cite{Wang:1991xy, Qin:2015srf, Blaizot:2015lma}. When high transverse momentum jet partons produced at the early stage of  heavy-ion collisions propagate through the hot and dense QGP, they interact with the medium constituents \cite{Gyulassy:1993hr,Baier:1994bd,Zakharov:1996fv,Gyulassy:2000fs,Wiedemann:2000za,Arnold:2001ba,Guo:2000nz}. This not only modifies the total energy of the jet partons, but also changes the distribution of the energy inside and outside the jet cone.

Earlier studies on jet-medium interaction and jet quenching in heavy-ion collisions have mainly concentrated on the nuclear suppression of high transverse momentum hadron production \cite{Adcox:2001jp,Adler:2002xw,Aamodt:2010jd,Abelev:2012hxa}, which tends to be more sensitive to the energy loss effect on the leading parton of jet \cite{Bass:2008rv,Armesto:2011ht,Burke:2013yra}.
The studies of jet-related correlation measurements, such as dihadron and $\gamma$-hadron correlations, have provided additional information on jet-medium interaction: the nuclear modification of the away-side jet or hadron yields may be sensitive to jet energy loss \cite{Zhang:2007ja, Zhang:2009rn, Qin:2009bk}, while the back-to-back angular distributions may be utilized to probe the medium-induced transverse momentum broadening \cite{Mueller:2016gko, Chen:2016vem, Chen:2016cof, Chen:2017zte}.

After the launch of the LHC, much attention has been paid to the production and nuclear modification of full jets in relativistic heavy-ion collisions \cite{Aad:2010bu,Chatrchyan:2011sx,Chatrchyan:2012gt,Chatrchyan:2012gw,Aad:2012vca,Chatrchyan:2012nia,Abelev:2013kqa,Chatrchyan:2013kwa, Aad:2014bxa,Aad:2014wha, Adam:2015doa}.
Various full jet observables have been explored and have provided much more detailed information about the interaction between jets and the QGP medium \cite{Vitev:2009rd,Qin:2010mn,CasalderreySolana:2010eh,Lokhtin:2011qq,Young:2011qx,He:2011pd,Renk:2012cx,Ma:2013pha,Senzel:2013dta,Chien:2015hda,Milhano:2015mng, Dai:2012am,Wang:2013cia,Qin:2012gp, MehtarTani:2011tz,CasalderreySolana:2012ef,Blaizot:2013hx,Fister:2014zxa, Apolinario:2012cg,Zapp:2012ak,Majumder:2013re,Casalderrey-Solana:2015vaa,Chang:2016gjp,Tachibana:2017syd}.
The nuclear modification of single inclusive jet rates, dijet and $\gamma$-jet transverse momentum imbalance distribution, etc., have clearly shown that full jets may experience a significant amount of energy loss during their propagation through the hot and dense QGP medium.
The nuclear modification of jet substructure observables, such as jet shape function and jet fragmentation function, have indicated that the inner hard cone of the full jets is difficult to be modified while the outer soft part of the jets may be easily affected by the traversed QGP.

Recently, a new jet substructure observable, namely, the momentum sharing ($z_g$) distribution of the groomed jets, has been studied in relativistic heavy-ion collisions \cite{Larkoski:2014wba,Larkoski:2015lea}.
It utilizes the jet grooming algorithms \cite{Butterworth:2008iy,Ellis:2009me,Krohn:2009th,Dasgupta:2013ihk,Larkoski:2014wba,Larkoski:2015lea} to investigate the internal structure of the full jets by removing the soft components of the jets.
In the soft drop de-clustering procedure \cite{Larkoski:2014wba,Larkoski:2015lea} as adopted by CMS and STAR Collaborations, a reconstructed full jet (with radius $R$ using the anti-$k_\mathrm{T}$ algorithm) is first re-clustered using the Cambridge-Aachen (C/A) algorithm and then de-clustered by dropping the softer branch until finding two hard branches with the following condition satisfied:
\begin{eqnarray}
\label{eq:zg}
\frac{\min(p_\mathrm{T1}, p_\mathrm{T2})}{p_\mathrm{T1}+p_\mathrm{T2}} \equiv z_g > z_\mathrm{cut}\left(\frac{\Delta R}{R} \right)^\beta,
\end{eqnarray}
where $p_\mathrm{T1}$ and $p_\mathrm{T2}$ are transverse momenta of the two hard sub-jets, $\Delta R$ is their angular separation, $z_\mathrm{cut}$ is the lower limit of the momentum sharing $z_g$ \cite{Larkoski:2015lea}.
CMS and STAR have measured the normalized $z_g$ distribution with $z_{\rm cut} = 0.1$, $\beta=0$ and $\Delta R \ge\Delta =0.1$ \cite{CMS:2016jys, Kauder:2017cvz}.
The momentum sharing $z_g$ distribution provides a unique opportunity to study the hard splitting of the partonic jet, and can be directly used to probe the medium-induced jet splitting function in the presence of the hot and dense QGP.

The experimental measurements from CMS Collaboration \cite{CMS:2016jys} have indeed seen strong nuclear modification of the groomed jet $z_g$ distribution at the LHC. Interestingly, the strength of the nuclear modification diminishes with increasing jet energies, which is consistent with some theoretical studies \cite{Chien:2016led,Mehtar-Tani:2016aco,KunnawalkamElayavalli:2017hxo}. However, the measurements from STAR Collaboration \cite{Kauder:2017cvz} have observed no obvious modification (at lower jet energies) at RHIC; this seems to contradict with the expectation from the CMS result in terms of the jet energy dependence of the nuclear modification of the $z_g$ distribution.

In this work, we present our study on the nuclear modification of the jet splitting in relativistic heavy-ion collisions at RHIC and the LHC based on the higher twist formalism \cite{Wang:2001ifa, Majumder:2009ge}.
We find a non-monotonic jet energy dependence of the nuclear modification of jet splitting function: strongest at intermediate jet energies, and weaker at larger or smaller jet energies.
This result is essential to explain the CMS-STAR groomed jet puzzle, i.e., strong nuclear modification of the momentum sharing $z_g$ distribution (with the strength diminishing with increasing jet $p_\mathrm{T}$) at the LHC whereas no obvious nuclear modification of the $z_g$ distribution at RHIC.
Another interesting finding is that the measured nuclear modification pattern of the $z_g$ distribution of groomed jets cannot be explained solely by independent energy loss of the two subjets.

The paper is organized as follows. In  Sec.~\ref{sec:framework} we provide a brief introduction to the framework that we utilize to study jet splitting and its nuclear modification in relativistic heavy-ion collisions.  In Sec.~\ref{sec:results} we present our numerical results and compare them to the experimental data of the groomed jet $z_g$ distribution from both CMS and STAR. In Sec.~\ref{sec:analysis} and \ref{sec:eLoss}, we analyze in detail the origin of the non-monotonic jet energy dependence of the nuclear modification of jet splitting function and the effect of independent energy loss for the two subjets. The summary is provided in Sec.~\ref{sec:summary}.

\section{Jet splitting in vacuum and QGP}
\label{sec:framework}

Using the jet grooming algorithms and the soft drop de-clustering procedure, CMS and STAR have measured the normalized distribution for the momentum sharing variable $z_g$ between the two subjets in the groomed jets,
\begin{eqnarray}
p(z_g) = \frac{1}{N_{\rm evt}}\frac{dN_{\rm evt}}{dz_g}.
\end{eqnarray}
The $z_g$ distribution may be directly related to the parton splitting function as follows \cite{Larkoski:2015lea,Chien:2016led}:
\begin{eqnarray}
\label{eq:pzg2}
p_i(z_g)=\frac{\int_{k_\Delta^2}^{k_R^2}dk_\perp^2\overline{P}_i(z_g,k_\perp^2)}{\int_{z_\mathrm{cut}}^{1/2}dx\int_{k_\Delta^2}^{k_R^2}dk_\perp^2\overline{P}_i(x,k_\perp^2)},
\end{eqnarray}
in which $i$ denotes the jet flavor, $k_\perp$ represents the transverse momentum of the splitted daughter parton with respect to the parent parton, $k_\Delta$ and $k_R$ are its lower and upper limits.
The symmetrized transverse momentum dependent parton splitting function $\overline{P}_i(x,k_\perp^2)$ is obtained by summing over all possible splitting channels:
\begin{eqnarray}
\overline{P}_i(x,k_\perp^2) = \sum_{j,l}\Big[{P}_{i\rightarrow j,l}(x,k_\perp^2)+{P}_{i\rightarrow j,l}(1-x,k_\perp^2)\Big]. \ \ \
\label{eq:partonSplit}
\end{eqnarray}
In the light cone coordinate, one may define the four-momentum of the parent parton at mid-rapidity as $[p_\mathrm{T},0,\vec{0}]$, and then the four-momenta of two daughter partons can be written as $[xp_\mathrm{T},{k_\perp^2}/({2xp_\mathrm{T}}),\vec{k}_\perp]$ and  $[(1-x)p_\mathrm{T},{k_\perp^2}/({2(1-x)p_\mathrm{T}}),-\vec{k}_\perp]$. One may derive the geometric relation between $k_\perp$ and the relative angle $\theta$ between the two daughter partons as: $k_\perp = 2p_\mathrm{T}x(1-x)\tan(\theta/2)$. Therefore, the boundaries of $k_\perp$ integration read: $k_\Delta = 2p_\mathrm{T}x(1-x)\tan(\Delta/2)$ and $k_R = 2p_\mathrm{T}x(1-x)\tan(R/2)$ in Eq.~(\ref{eq:pzg2}).

In the presence of QGP medium, the parton splitting function in Eq.~(\ref{eq:partonSplit}) has both vacuum and medium-induced contributions:
\begin{equation}
{P}_i(x, k_\perp^2)={P}^\mathrm{vac}_i(x, k_\perp^2)+{P}^\mathrm{med}_i(x, k_\perp^2).
\label{eq:vacmed}
\end{equation}
The vacuum part of the splitting function reads:
\begin{eqnarray}
	{P}^\mathrm{vac}_i(x, k_\perp^2)=\frac{\alpha_s}{2\pi}{P}^\mathrm{vac}_i(x)\frac{1}{k_\perp^2},
\label{eq:vacP}
\end{eqnarray}
where $\alpha_s$ is the strong coupling constant and ${P}^\mathrm{vac}_i(x)$ is the standard transverse momentum independent splitting function in vacuum.
To account for the running coupling effect, we take the scale in $\alpha_s$ as: $Q^2 = k_\perp^2/[x(1-x)]$.
In this work, the medium-induced contribution to the parton splitting function is taken from the higher twist formalism \cite{Wang:2001ifa, Majumder:2009ge}:
\begin{eqnarray}
{P}^\mathrm{med}_i(x,k_\perp^2)=\frac{2 \alpha_s}{\pi k_\perp^4}{P}^\mathrm{vac}_i(x) \int d\tau \hat{q}_g(\tau)\sin^2\left(\frac{\tau}{2\tau_f}\right), \ \ \
\label{eq:medP}
\end{eqnarray}
where $\tau$ represents the time of parton-medium interaction, $\hat{q}_g$ is the gluon jet transport coefficient which denotes the transverse momentum broadening per unit time via elastic scatterings, and $\tau_f=2Ex(1-x)/k^2_\perp$ is the formation time of the radiation (splitting) with $E$ as the energy of the parent parton.

To evaluate the medium-induced splitting function in realistic heavy-ion collisions, we couple Eq.~(\ref{eq:medP}) to hydrodynamic models \cite{Pang:2012he,Pang:2013pma,Song:2007ux,Qiu:2011hf} that provide the space-time evolution of the QGP fireballs. With the knowledge of the local temperature $T$ and flow four-velocity $u$ along the path of a given jet, the local $\hat{q}$ is calculated via:
\begin{equation}
\label{eq:qhat}
\hat{q} (\tau,\vec{r}) = \hat{q}_0 \frac{T^3(\tau,\vec{r})}{T_{0}^3(\tau_{0},\vec{0})} \frac{p\cdot u(\tau, \vec{r})}{p^0},
\end{equation}
where $T_{0}(\tau_{0},\vec{0})$ and $\hat{q}_0$ are defined as the initial temperature and transport coefficient at the center of central (0-10\%) A+A collisions, $p^\mu$ is the four-momentum of the propagating parton, and the ${p\cdot u}/{p^0}$ factor takes into account the flow effect on the effective value of $\hat{q}$ \cite{Baier:2006pt}.
Throughout our following discussion, $\hat{q}$ generally denotes the quark transport coefficient; the gluon transport coefficient $\hat{q}_g$ can be converted via the Casmir factors $\hat{q}C_A/C_F$.

To directly compare to the measured jet splitting function $p^\mathrm{obs}(z_g)$ observed at a given $p_\mathrm{T} \in (p_\mathrm{T,1}, p_\mathrm{T,2})$ range in the experiments, one needs to convolute the above splitting function $p(z_g)$ with both the initial parton spectra and the jet energy loss calculation (in the presence of the medium).
The expression for the observed $z_g$ distribution reads:
\begin{align}
p^\mathrm{obs}(z_g)&=\frac{1}{\sigma_{\rm total}}\sum_{j=q,g}\int d^2X \mathcal{P}(\vec{X}) \nonumber\\
&\times  \int_{p_\mathrm{T,1}^{\rm ini} = p_\mathrm{T,1}^{\rm obs}+\Delta E_1}^{p_\mathrm{T,2}^{\rm ini} = p_\mathrm{T,2}^{\rm obs}+\Delta E_2} dp_\mathrm{T}^{\rm ini} \frac{d\sigma_{j}}{dp_\mathrm{T}^{\rm ini}} p_j(z_g|p_\mathrm{T}^{\rm ini}).
\label{pzgX2}
\end{align}
Here, $\mathcal{P}(\vec{X})$ denotes the probability distribution for the initial production vertex $\vec{X}$ of the jet parton in the transverse plane (for A+A collisions) which is obtained via the Glauber model calculation.
The initial momentum space distribution of jet partons are calculated using the leading order perturbative QCD cross sections convoluted with the CTEQ parameterizations \cite{Lai:1999wy} for the parton distribution function and the EPS09 parametrizations \cite{Eskola:2009uj} for the initial state nuclear shadowing effect in A+A collisions.
In the following discussion, the superscript ``obs" on the left hand side of Eq.~(\ref{pzgX2}) will be neglected for simplicity.
The energy loss $\Delta{E}$ experienced by a given jet, which depends on the path (and thus its production vertex and propagation direction), can in principle be obtained via the integral of the medium-induced splitting function weighted by the energy of emitted gluon:
\begin{equation}
\label{eq:eloss}
	\Delta{E} = \int dx dk^2_\perp (xE) \overline{P}^\mathrm{med}(x,k^2_\perp) \theta(\frac{1}{2}-x) \theta(k_\perp-k_R).
\end{equation}
Here, we only consider the energy loss due to the out-of-cone radiation and treat the energy loss of the two splitted subjets as a single parent jet parton interacting with the medium and lose energy coherently.
This is motivated by the fact that using independent energy loss for the two subjets cannot explain the observed pattern for the nuclear modification of the momentum sharing $z_g$ distribution, as will be illustrated in details in Sec.~\ref{sec:eLoss}.
Note that the semi-analytical evaluation of Eq.~(\ref{eq:eloss}) neglects the effects due to the possible multiple soft splittings and the variation of parton energy during the propagation through the medium.
To account for these effects, in this work we use the linear Boltzmann transport (LBT) model \cite{Cao:2016gvr,Cao:2017hhk,Chen:2017zte}, that is based on the same higher twist formalism, to simulate the medium modified parton showers through the QGP and extract the amount of energy ($\Delta E$) that flows outside the jet cone $R$.
Since a large cone size $R=0.4$ is used in this work, $\Delta E$ is usually small as compared to the initial jet energy.
We have verified that the obtained jet energy loss with this state-of-the-art parton-by-parton simulation quantitatively agrees with the direct estimation using Eq.~(\ref{eq:eloss}) except for very small jet energies for which the effect of jet energy variation during its propagation is more important.

\section{Nuclear modification of jet splitting at RHIC and the LHC}
\label{sec:results}

In this section, we provide the numerical results for the nuclear modification of jet splitting function at the LHC and RHIC and compare them to the experimental data from CMS and STAR Collaborations. The nuclear modification factor of jet splitting function, $R_{p(z_g)}$, is defined as the ratio between the medium-modified $p(z_g)$ in A+A collisions and the vacuum $p(z_g)$ in p+p collisions:
\begin{eqnarray}
R_{p(z_g)}=\frac{p(z_g)\mid_\mathrm{AA}}{p(z_g)\mid_\mathrm{pp}},
\label{ratio_pzg}
\end{eqnarray}
where the denominator is the p+p baseline that only includes the vacuum contribution, while the numerator is for A+A collisions and includes both vacuum and medium-induced contributions.

\begin{figure}[tb]
	\includegraphics[width=0.99\linewidth]{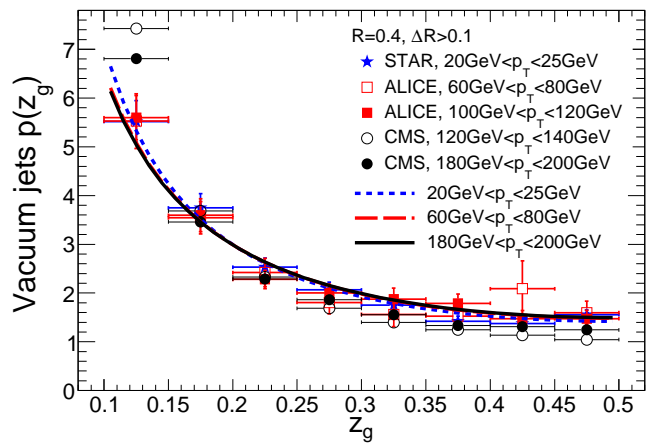}
	\caption{(Color online) Vacuum jet splitting function for different jet $p_\mathrm{T}$ ranges compared to the groomed jet measurements in p+p collisions at $\sqrt{s_{NN}} = 5.02$~TeV by CMS \cite{CMS:2016jys}, in p+p collisions at $\sqrt{s_{NN}} = 200$~GeV by STAR \cite{Kauder:2017cvz}, and in p-Pb collisions at $\sqrt{s_{NN}} = 5.02$~TeV by ALICE \cite{Caffarri:2017bmh}. }
  \label{fig:p_zg_vacuum}
\end{figure}

\begin{figure}[tb]
\includegraphics[width=0.99\linewidth]{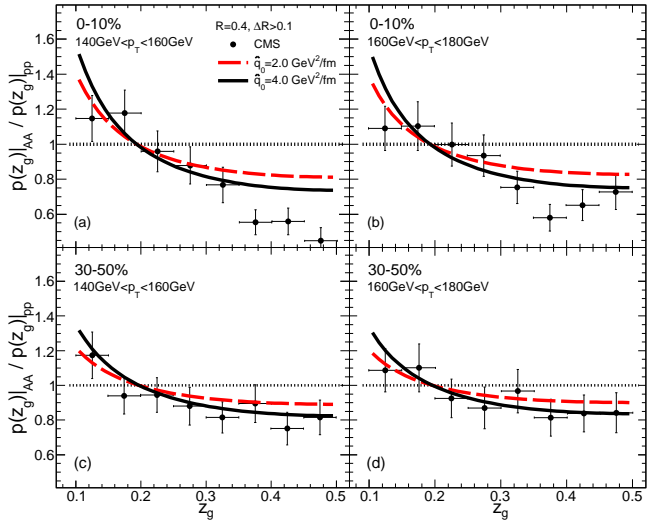}
  \caption{(Color online) Nuclear modification factor $R_{p(z_g)}$ of jet splitting function in central 0-10\% and mid-peripheral 30-50\% Pb+Pb collisions at 5.02~ATeV for two different jet $p_\mathrm{T}$ ranges.
	The experimental data are taken from CMS \cite{CMS:2016jys}.}
  \label{fig:ratio_pzg_centrality}
\end{figure}

\begin{figure}[tb]
\includegraphics[width=0.99\linewidth]{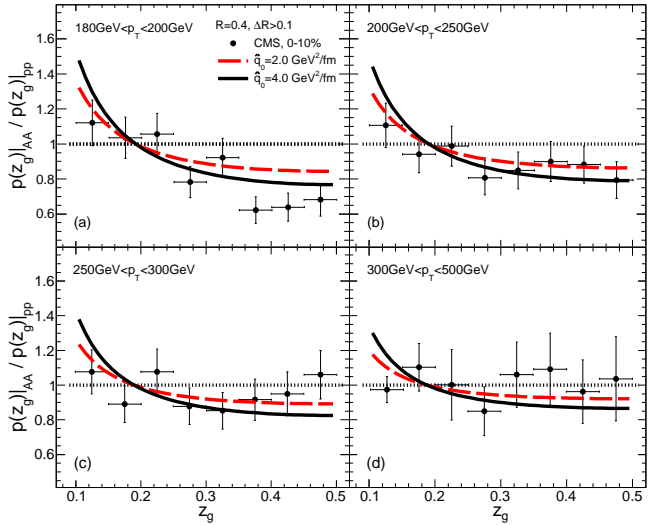}
  \caption{(Color online) Nuclear modification factor $R_{p(z_g)}$ of jet splitting function for four different jet $p_\mathrm{T}$ ranges in central 0-10\% Pb+Pb collisions at 5.02~ATeV. The experimental data are taken from CMS \cite{CMS:2016jys}.}
  \label{fig:ratio_pzg_high_pT}
\end{figure}

\begin{figure}[tb]
\includegraphics[width=0.99\linewidth]{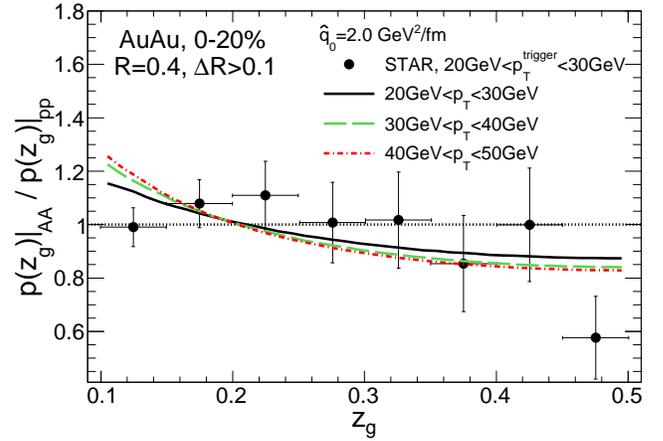}
	\caption{(Color online) Nuclear modification factor $R_{p(z_g)}$ of single inclusive jet splitting function in central 0-20\% Au+Au collisions at 200~AGeV. The experimental data are taken for the triggered jets in dijet events from STAR \cite{Kauder:2017cvz}. }
  \label{fig:ratio_zg_e_qhat2_rhic}
\end{figure}

We first show in Fig. \ref{fig:p_zg_vacuum} the vacuum jet splitting function for different jet $p_\mathrm{T}$ ranges.
The experimental data for the groomed jet momentum sharing $z_g$ distribution in p+p collisions at $\sqrt{s_{NN}} = 5.02$~TeV from CMS Collaboration, in p+p collisions at $\sqrt{s_{NN}} = 200$~GeV from STAR Collaboration, and in p-Pb collisions at $\sqrt{s_{NN}} = 5.02$~TeV from ALICE Collaboration, are shown for comparison.
Our calculation shows very weak dependence on the jet $p_\mathrm{T}$ of the vacuum jet splitting function.
One can see that a reasonable description of the $z_g$ distribution in p+p (and p-Pb) collisions is obtained from our calculation, which serves as the baseline for studying the medium modification in relativistic heavy-ion collisions.

In Fig.~\ref{fig:ratio_pzg_centrality}, we present the nuclear modification factor $R_{p(z_g)}$ of single inclusive jet splitting function in central (0-10\%) and mid-peripheral (30-50\%) Pb+Pb collisions at $5.02$~ATeV. The evolution profile of the QGP medium is provided by the (3+1)-dimensional  hydrodynamic model CLVisc \cite{Pang:2012he,Pang:2013pma}.
Figure~\ref{fig:ratio_pzg_centrality} shows the results using two different values of $\hat{q}_0$ (2 and 4~GeV$^2$/fm).
One can see that the modification of jet splitting function is stronger for larger value of $\hat{q}_0$ and in more central collisions.
The use of $\hat{q}_0=4$~GeV$^2$/fm can describe quite well the nuclear modification data of the momentum sharing $z_g$ distribution from CMS Collaboration.
To further investigate the transverse momentum dependence of $R_{p(z_g)}$, we present in Fig.~\ref{fig:ratio_pzg_high_pT} the nuclear modification factor of jet splitting function for more $p_\mathrm{T}$ ranges in central (0-10\%) Pb+Pb collisions.
Combining with the two panels for central (0-10\%) collisions in Fig.~\ref{fig:ratio_pzg_centrality}, we observe that within the $p_\mathrm{T}$ ranges explored here (and by CMS), the nuclear modification becomes weaker with increasing jet $p_\mathrm{T}$.

In Fig.~\ref{fig:ratio_zg_e_qhat2_rhic}, we show the nuclear modification factor $R_{p(z_g)}$ of single inclusive jet splitting function in central (0-20\%) Au+Au collisions at $200$~AGeV and compare to the STAR result for the triggered jets in dijet events (the result for the recoiled jets from STAR is quantitatively similar) \cite{Kauder:2017cvz}.
Here, the QGP medium is simulated utilizing the (2+1)-dimensional viscous hydrodynamic model VISHNew \cite{Song:2007ux, Qiu:2011hf}.
The initial quark transport coefficient is taken to be $\hat{q}_0$ as 2~GeV$^2$/fm in central Au+Au collisions at 200~AGeV, which is half of that for central Pb+Pb collisions at 5.02~ATeV due to the lower initial density and temperature of the QGP medium.
We see that our calculation provides a good description of the STAR triggered jet result for the nuclear modification of the groomed jet $z_g$ distribution as well.

One interesting feature from the RHIC result is that both STAR data and our calculation show much smaller nuclear modification of the jet splitting function at RHIC than at the LHC.
One obvious effect is the smaller $\hat{q}_0$ and medium size due to the lower collision energy (thus lower initial energy density and temperature) at RHIC compared to the LHC.
But even with the same value of $\hat{q}_0 = 2$~GeV$^2$/fm for both RHIC and the LHC, the nuclear modification is still weaker at RHIC, which is hard to explain based on the naive expectation from the LHC result, i.e., the nuclear modification is stronger for smaller jet energies (thus ``should" be not small at RHIC).
The origin of this non-monotonic jet energy dependence of the nuclear modification of jet splitting function will be analyzed in detail in the next section.

\section{Non-monotonic jet energy dependence of $R_{p(z_g)}$}
\label{sec:analysis}

In the previous section, we have seen that the nuclear modification of jet splitting function exhibits a non-monototic dependence on jet energies.
As we will show in this section, such non-monototic behavior originates from two competing factors when jet $p_\mathrm{T}$ increases (decreases): (1) the medium-induced contribution to jet splitting function as compared to the vacuum contribution becomes smaller (larger); (2) the shape of the medium-induced splitting function with respect to $x$ or $z_g$ becomes deeper (flatter). The first factor can explain the diminishing nuclear modification of the groomed jet momentum sharing $z_g$ distribution with increasing jet $p_\mathrm{T}$ (observed by CMS), while the second factor can explain the observation of the small modification of the groomed jet $z_g$ distribution for lower $p_\mathrm{T}$ jets at RHIC (observed by STAR). The combination of the above two effects generates a non-monotonic jet $p_\mathrm{T}$ dependence of the nuclear modification of jet splitting function.

The first factor can be easily understood from the fact that within the higher twist formalism, the medium-induced splitting function is directly controlled by $\int d\tau \hat{q}(\tau)/k_{\perp}^2$ as compared to the vacuum splitting function. This implies that for a given jet cone size $R$, since $k_\perp$ is directly related to jet energy, the medium-induced contribution to jet splitting function and thus the nuclear modification of the momentum sharing $z_g$ distribution tends to diminish for increasing jet energies.
To quantitatively illustrate such effect, we may define the fractional contribution $F_i^m$ from the medium-induced splitting to the integrated jet splitting probability as follows:
\begin{eqnarray}
F_i^m = \frac{\int_{z_\mathrm{cut}}^{1/2}dx\int_{k_\Delta^2}^{k_R^2}dk^2_\perp\overline{P}_i^\mathrm{med}(x,k^2_\perp)}{\int_{z_\mathrm{cut}}^{1/2}dx\int_{k_\Delta^2}^{k_R^2}dk_\perp^2 \left[\overline{P}_i^\mathrm{vac}(x,k_\perp^2) + \overline{P}_i^\mathrm{med}(x,k_\perp^2)\right]}.\nonumber\\
\label{r_int_pzg}
\end{eqnarray}

\begin{figure}[tb]
\includegraphics[width=0.99\linewidth]{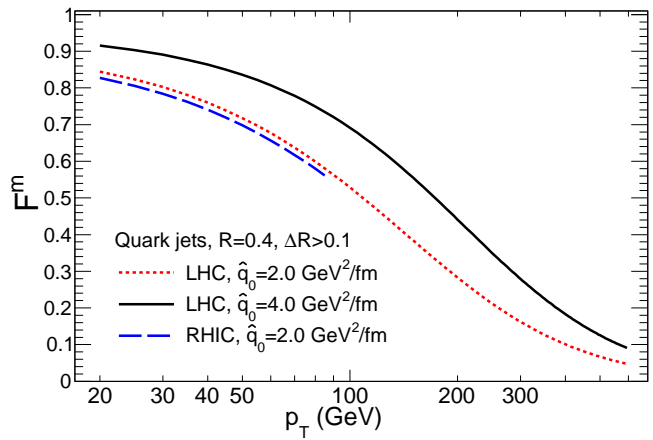}
   \caption{(Color online) The fractional contribution $F^m$ from medium-induced jet splitting to the integrated jet splitting probability as a function of jet $p_\mathrm{T}$ at RHIC and the LHC.}
  \label{fig:p_zg_sum_ratio}
\end{figure}

\begin{figure}[tb]
\includegraphics[width=0.99\linewidth]{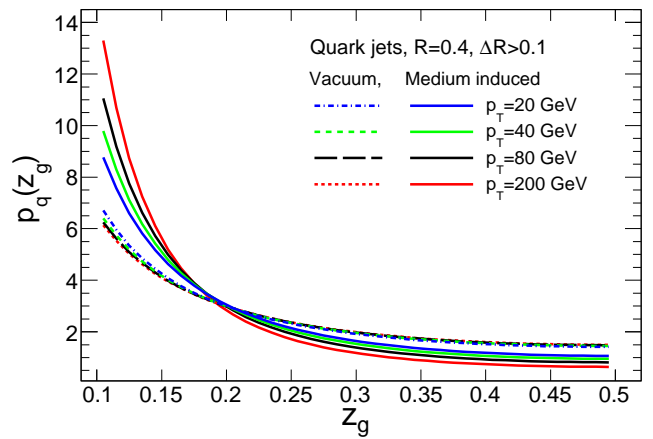}
	\caption{(Color online) Vacuum and medium-induced $z_g$ distribution for different values of jet $p_\mathrm{T}$ at the LHC.}
  \label{fig:zg_comp_induced}
\end{figure}

\begin{figure}[tb]
\includegraphics[width=0.99\linewidth]{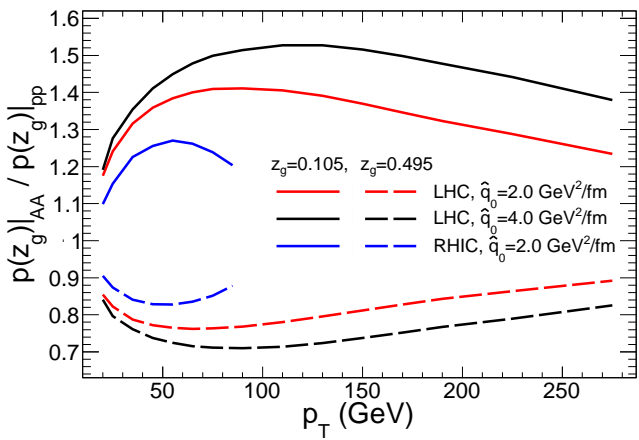}
	\caption{(Color online) Jet $p_\mathrm{T}$ dependence of the nuclear modification factor $R_{p(z_g)}$ of the momentum sharing distribution for groomed jets at RHIC and the LHC for two $z_g$ values: $z_g=0.105$ and $z_g=0.495$.}
  \label{fig:ratio_pzg_e_evo_rhic-lhc3}
\end{figure}

In Fig.~\ref{fig:p_zg_sum_ratio}, we show this fractional contribution $F_i^m$ for quark jets as a function of the jet $p_\mathrm{T}$.
One observes that the contribution from the medium-induced splitting to the total jet splitting probability becomes smaller as one increases the jet energies at both RHIC and the LHC.
The medium-induced contribution becomes very small for jet $p_\mathrm{T}>400$-$500$~GeV at the LHC.
This explains the observation by CMS Collaboration that the nuclear modification of the groomed jet momentum sharing $z_g$ distribution diminishes for very large jet $p_\mathrm{T}$.

On the other hand, as jet $p_\mathrm{T}$ decreases, the contribution from the medium-induced splitting compared to the vacuum splitting becomes more important.
However, the nuclear modification of the normalized momentum sharing $z_g$ distribution also depends on the shape of the medium-induced splitting function.
In fact, the shape of the medium-induced splitting function with respect to the momentum fraction $x$ or $z_g$ are different between large and small jet energies.
Within the higher twist formalism, one can show that:
\begin{eqnarray}
   \int_{k_\Delta^2}^{k_R^2}dk_\perp^2 \overline{P}_i^\mathrm{med}(x,k_\perp^2) \rightarrow \begin{cases}
1/x\,, & \mathrm{small}\:\, E; \\
1/x^3\,, & \mathrm{large}\:\, E. \\
\end{cases}
\end{eqnarray}
This indicates that with decreasing jet energy, the shape of the medium-induced splitting function with respect to $x$ becomes flatter, changing from $1/x^3$ to $1/x$ for large and small jet energy limits.
More quantitative result for this behavior is shown in Fig. \ref{fig:zg_comp_induced}.
One can see that the medium-induced splitting function becomes flatter with decreasing jet $p_\mathrm{T}$.
Combining with the smaller $\hat{q}_0$ and smaller size of QGP medium at RHIC than at the LHC, our calculation can explain why STAR observes  little nuclear modification of the momentum sharing $z_g$ distribution for the groomed jets with $p_\mathrm{T}=10$-$30$~GeV.

Since there exists a non-monotonic dependence on jet energy, the maximal nuclear modification of the jet splitting function should be in the intermediate jet $p_\mathrm{T}$ regime.
In Fig. \ref{fig:ratio_pzg_e_evo_rhic-lhc3}, we show our prediction for the nuclear modification of the groomed jet momentum sharing distribution as a function of jet $p_\mathrm{T}$.
For the purpose of good resolution, we only show the values of the nuclear modification factor $R_{p(z_g)}$ around the two endpoints: $z_g = 0.1$ and $z_g = 0.5$.
The non-monotonic jet energy dependence of $R_{p(z_g)}$ can be clearly seen for both RHIC and the LHC.
Note that the nuclear modification effect is smaller at RHIC than at the LHC (even with the same value of $\hat{q}_0$), which mainly originates from the fact that both the size and the initial density of the QGP medium are smaller at RHIC than at the LHC.
From Fig. \ref{fig:ratio_pzg_e_evo_rhic-lhc3}, we see that the maximal nuclear modification of the groomed jet momentum sharing $z_g$ distribution is at around $50$-$60$~GeV at RHIC and at around $70$-$90$~GeV at the LHC, respectively.
Future measurements of groomed jets with a wider range of jet $p_\mathrm{T}$ (lower jet energies at the LHC or larger jet energies at RHIC) should be able to test our result.

\section{Effect of independent subjet energy loss}
\label{sec:eLoss}

In the previous sections, we have calculated the energy loss of jets due to the out-of-cone radiation assuming that the two subjets interact with the QGP medium coherently and lose energy like a single parent parton. Based on such coherent energy loss assumption for the two subjets, we are able to explain the CMS-STAR puzzle and have found a non-monotonic jet energy dependence of the nuclear modification of jet splitting function.
In this section, we will show that the application of independent (incoherent) energy loss to the two splitted subjets cannot explain the nuclear modification pattern of the groomed jet momentum sharing $z_g$ distribution observed by CMS and STAR Collaborations.

Consider a jet splitting into two subjets with the momentum fractions $z_1$ and $z_2 = 1-z_1$. Here we take $z_1\le 0.5\le z_2$ and thus the initial momentum sharing variable $z_g^{\rm ini}=z_1$.
If the two subjets lose energy in the QGP medium independently, then after traversing the QGP, their energy fractions with respect to the initial jet energy will be changed and are denoted as $z_1'$ and $z_2'$.
From $z_1/z_2\le 1$, one typically obtains $z_1'/z_2'\le z_1/z_2\le 1$ since the fractional energy loss usually decreases with increasing jet energy.
One may further show that the final momentum sharing variable $z_g^{\rm fin} = z_1'/(z_1'+z_2') \le z_g^{\rm ini}$; this implies the momentum sharing $z_g$ becomes smaller after independent energy loss of the two subjets.

\begin{figure}[tb]
	\includegraphics[width=0.99\linewidth]{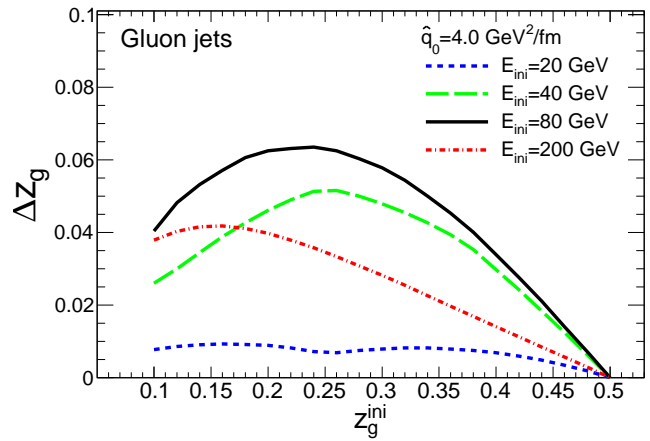}
	\caption{(Color online) The relation between $\Delta z_g$ and $z_g^{\rm ini}$ considering independent energy loss of the two subjets originated from gluon jets (at the LHC).
  }
  \label{fig:zgini_zgfinal}
\end{figure}

We can test the above effect using the linear Boltzmann transport model \cite{Cao:2016gvr,Cao:2017hhk,Chen:2017zte}.
Here for simplicity, we assume that the jet splits into two subjets at the starting time of the QGP formation and then evolve in the QGP independently.
The numerical result for the shift of the $z_g$ value due to the independent energy loss of the two subjets is shown in Fig. \ref{fig:zgini_zgfinal} for various jet $p_\mathrm{T}$.
We observe that the value of the $z_g$ variable indeed becomes smaller, which means the energies carried by the two subjets become more asymmetric after independent energy loss.
One naive expectation would be that the momentum sharing variable $z_g$ distribution becomes deeper.
However, this is not the case.
Note that the value of $z_g^{\rm fin}$ can be smaller than the cut $z_{\rm cut}=0.1$ after independent energy loss of the two subjets, as shown in Fig. \ref{fig:zgini_zgfinal}.
Events with $z_g<z_{\rm cut}=0.1$ do not contribute to the final normalized $z_g$ distribution, therefore, the above expectation does not hold.

\begin{figure}[tb]
\includegraphics[width=0.99\linewidth]{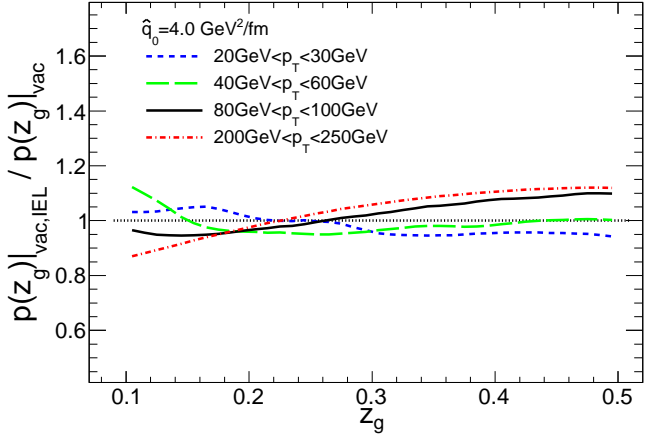}
	\caption{(Color online) The modification effect on the vacuum $z_g$ distribution at the LHC due to the $z_g$ shift caused by independent energy loss (IEL) for subjets.
  }
  \label{fig:ratio_pzg_e_to_vacuum}
\end{figure}

To perform a more quantitative analysis, we incorporate the effect of independent subjet energy loss on the modification of the groomed jet $z_g$ distribution as:
\begin{eqnarray}
\frac{dN^{\rm fin}}{dz_g^{\rm fin}} = \int d\Delta z_g P(\Delta z_g|z_g^{\rm ini}) \left.\frac{dN^{\rm ini}}{dz_g^{\rm ini}}\right|_{z_g^{\rm ini}=z_g^{\rm fin}+\Delta z_g},
\label{eq:pzg_shift}
\end{eqnarray}
where $P(\Delta z_g|z_g^{\rm ini})$ accounts for the effect of the momentum sharing variable shift $\Delta z_g$ due to the independent subjet energy loss.
Since after subjet energy loss, the value of $z_g^{\rm fin}$ might be smaller than $z_{\rm cut}=0.1$, the final normalized momentum sharing distribution should be obtained as:
\begin{eqnarray}
p^{\rm fin}(z_g^{\rm fin}) = \frac{ \frac{dN^{\rm fin}}{dz_g^{\rm fin}}}{\int_{z_{\rm cut}}^{1/2} dz_g^{\rm fin} \frac{dN^{\rm fin}}{dz_g^{\rm fin}}}.
\end{eqnarray}

First, we apply the independent subjet energy loss effect on the vacuum splitting function; this mimics the picture that the vacuum-splitted subjets lose energy independently in the QGP medium via medium-induced radiative process.
Similar study has also been performed in Ref. \cite{Mehtar-Tani:2016aco}.
Our result for the nuclear modification factor $R_{p(z_g)}$ is shown in Fig. \ref{fig:ratio_pzg_e_to_vacuum}.
One can see that the modification pattern of momentum sharing variable $z_g$ distribution is very different from the experimental data on the nuclear modification of the normalized $z_g$ distribution obtained by CMS and STAR Collaborations.
In particular for large jet $p_\mathrm{T}$, there is an enhancement at large $z_g$ and a suppression at small $z_g$ for the normalized $z_g$ distribution.
This indicates independent subjet energy loss may flatten the normalized $z_g$ distribution of the groomed jets (in contradiction to the naive expectation from the $z_g$ shift to smaller values).

In order to understand more clearly the independent subjet energy loss effect on the $z_g$ distribution, we may analyze Eq.~(\ref{eq:pzg_shift}) by taking the initial $z_g$ distribution to be a power law, i.e., ${dN^{\rm ini} } / {dz_g^{\rm ini}} \sim 1/(z_g^{\rm ini})^\alpha$.
Using a $\delta$ function for the probability distribution $P(\Delta z_g|z_g^{\rm ini})$, one may perform the integration over $d\Delta z_g$.
The Taylor expansion for small $\Delta z_g$ renders:
\begin{eqnarray}
	\frac{dN^{\rm fin}}{dz_g^{\rm fin}} \approx \frac{1}{|\mathcal{J}|} \frac{1}{(z_g^{\rm fin})^\alpha} \left({ 1-\alpha \frac{\Delta z_g}{z_g^{\rm fin}} + \cdots }\right), \ \ \
\end{eqnarray}
where $\mathcal{J} = { 1 - \left.\frac{d\Delta z_g}{dz_g^{\rm ini}}\right|_{z_g^{\rm ini}=z_g^{\rm fin}+\Delta z_g}}$ is the Jacobian factor.
One can see from the above equation that the effect of independent subjet energy loss on the normalized $z_g$ distribution is controlled by both the shape of the initial $z_g$ distribution and the details of the $z_g$ shift ($\Delta z_g$) as a function of $z_g^{\rm ini}$.

Applying the above argument to Fig. \ref{fig:ratio_pzg_e_to_vacuum}, since the initial input is the vacuum splitting function with a roughly fixed $\alpha$ value [$p_{\rm vac}(z_g) \sim 1/z_g$, $\alpha \sim 1$], the nuclear modification pattern of the normalized $z_g$ distribution is almost entirely determined by the shift $\Delta z_g$ as a function of $z_g$.
From Fig. \ref{fig:zgini_zgfinal}, one can see that the relative $z_g$ shift (the value of $\Delta z_g/z_g$) at large $z_g$ is typically smaller than that at small $z_g$.
Therefore, the shape of the $z_g$ distribution will indeed become flatter after the inclusion of the $z_g$ shift caused by independent subjet energy loss.
In addition to such $z_g$ shift effect, the Jacobian factor $|\mathcal{J}|^{-1}$ also plays a role in explaining the detailed modification pattern as shown in Fig. \ref{fig:ratio_pzg_e_to_vacuum}.
From Fig. \ref{fig:zgini_zgfinal}, we can see that $d\Delta z_g/dz_g$ is positive ($|\mathcal{J}|^{-1}>1$) for small $z_g$ and negative ($|\mathcal{J}|^{-1}<1$)  for large $z_g$.
This means that the Jacobian factor will cause the suppression effect at large $z_g$ and the enhancement at low $z_g$; this effect is opposite to the relative $z_g$ shift.
The combination of the above two competing effects gives the rich modification pattern in Fig. \ref{fig:ratio_pzg_e_to_vacuum} for the normalized $z_g$ distribution caused by independent subjet energy loss.

\begin{figure}[tb]
\includegraphics[width=0.99\linewidth]{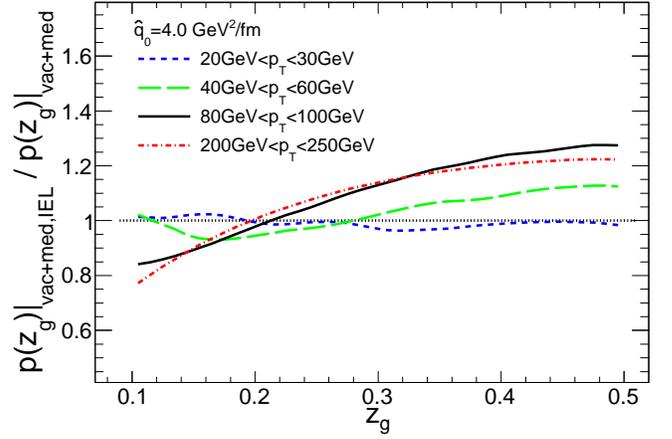}
	\caption{(Color online) The modification effect on the medium-modified $z_g$ distribution at the LHC due to the $z_g$ shift caused by independent energy loss (IEL) for subjets.
  }
  \label{fig:pzg_e_zg_shift}
\end{figure}

\begin{figure}[tb]
\includegraphics[width=0.99\linewidth]{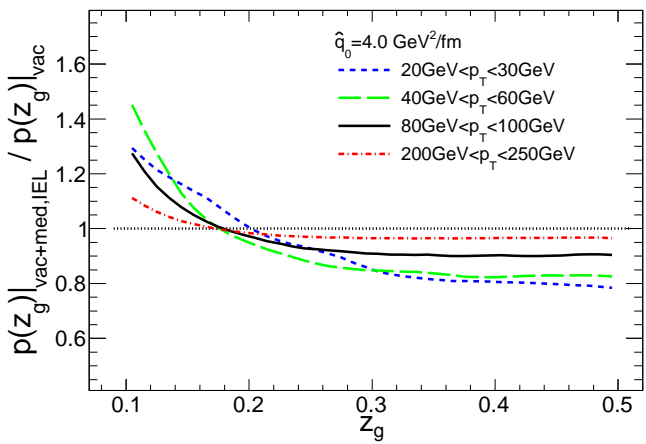}
	\caption{(Color online) The nuclear modification factor $R_{p(z_g)}$ of the groomed jet momentum sharing $z_g$ distributions at the LHC after the inclusion of the effects from medium-induced splitting and the independent energy loss (IEL) for subjets.
  }
  \label{fig:ratio_pzg_e_zg-shift_ratio}
\end{figure}

The above flattening effect on the normalized $z_g$ distribution caused by the $z_g$ shift also depends the initial $z_g$ distribution (i.e., the value of the power index $\alpha$): the flattening effect is typically stronger for larger values of $\alpha$.
This can be clearly seen in Fig. \ref{fig:pzg_e_zg_shift}, where we show the effect of independent subjet energy loss applied on the medium-modified $z_g$ distribution.
This mimics the picture that the two subjets produced from medium-modified splitting lose energy independently in the QGP medium.
Compared to Fig. \ref{fig:ratio_pzg_e_to_vacuum}, we indeed observe a stronger flattening effect on the normalized $z_g$ distribution; this is because with the inclusion of the medium-induced splitting contribution, the initial $z_g$ distribution used here is deeper than the vacuum $z_g$ distribution.

In Fig. \ref{fig:ratio_pzg_e_zg-shift_ratio}, we show the nuclear modification factor $R_{p(z_g)}$ of the groomed jet $z_g$ distribution with the inclusion of independent subjet energy loss as well as the medium-induced splitting.
We see that the medium-modified normalized $z_g$ distribution as compared to the vacuum one is suppressed at large $z_g$; such effect is larger for smaller jet $p_\mathrm{T}$.
Compared to the previous section where the coherent energy loss of the two subjets is applied, the most interesting feature is that for a wide range of jet $p_\mathrm{T}$ explored here, the non-monotonic jet $p_\mathrm{T}$ dependence of the nuclear modification factor $R_{p(z_g)}$ disappears when the independent energy loss of the two subjets is applied together with the inclusion of the medium-induced splitting contribution.
This seems to suggest that the independent (incoherent) subjet energy loss cannot explain the experimental data on the nuclear modification factor of the normalized $z_g$ distribution obtained by CMS and STAR Collaborations.
Considering that the two subjets would tend to decouple from each other for sufficiently large angular separation, it would be interesting to vary the angular separation between the two subjets in future measurements in order to explore coherent and independent subjet energy loss scenarios and in-between.

\begin{figure}[tb]
\includegraphics[width=0.99\linewidth]{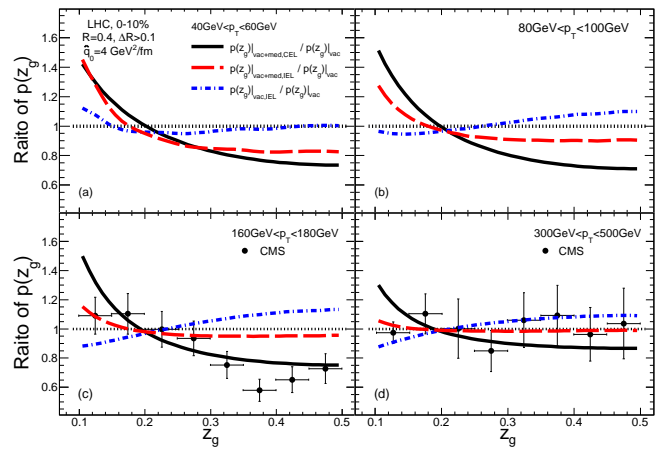}
	\caption{(Color online) Modification of the groomed jet $z_g$ distributions at the LHC for three different scenarios: solid for medium-modified splitting with coherent energy loss (CEL) of subjets, dashed for medium-modified splitting with independent energy loss (IEL) of subjets, and dash-dotted for vacuum splitting with IEL of subjets.
  }
  \label{fig:ratio_pzg_e_4bins_3-scenarios}
\end{figure}

Finally, to illustrate the effects of different jet-medium interaction mechanisms on the modification of the $z_g$ distribution, we show in Fig. \ref{fig:ratio_pzg_e_4bins_3-scenarios} the modification patterns of the groomed jet momentum sharing $z_g$ distributions for three different scenarios.
For Case (1), we apply coherent energy loss of subjets on the medium-modified $z_g$ distribution, which has been described in details in Sections II, III and IV.
For Case (2), independent energy loss of subjets is applied on the medium-modified $z_g$ distribution; this describes the situation that the two subjets produced from medium-modified splitting lose energy independently in the QGP medium.
For Case (3), we put independent subjet energy loss effect on the vacuum splitting function, which mimics the picture that the vacuum-splitted subjets lose energy independently via the medium-induced radiative process.
Case (2) and Case (3) have been elaborated above in this section.
One can see the clear difference among different jet-medium interaction scenarios, and only Case (1) can describe the CMS (and STAR) groomed jet measurements, especially the jet $p_{\mathrm T}$ dependence for the nuclear modification of the momentum sharing $z_g$ distribution.

\section{Summary}
\label{sec:summary}

We have studied the nuclear modification of jet splitting in relativistic heavy-ion collisions at RHIC and the LHC energies based on the higher twist formalism.
It is interesting to find that different subjet energy loss scenarios produce different nuclear modification patterns of jet splitting function.
Our result shows that the observed nuclear modification pattern of the $z_g$ distribution of groomed jets cannot be explained solely by the independent (incoherent) energy loss of the two splitted subjets.
In contrast, with the assumption of coherent energy loss of the two subjets in the QGP medium, we have found a non-monotonic jet energy dependence of the nuclear modification of jet splitting function: the maximal modification at intermediate jet energies and diminishing modification at larger and smaller jet energies.
Combined with the smaller size and lower density of the QGP medium at RHIC than at the LHC, our result can explain the current puzzle between CMS and STAR groomed jet measurements: strong nuclear modification has been observed for the momentum sharing $z_g$ distribution at the LHC while no obvious modification of the $z_g$ distribution has been seen at RHIC.

This work constitutes an important contribution to the study of jet-medium interaction and jet substructure in relativistic heavy-ion collisions.
Future measurements of the groomed jets with a wider range of jet energies at both the LHC and RHIC can test our finding about the non-monotonic jet energy dependence of the nuclear modification of jet splitting function.
The groomed jet measurements with varying angular separation between the two subjets can also provide more detailed information and more stringent constraint on our understanding of coherent and independent energy loss scenarios for full jet shower evolution in the hot and dense QGP produced in high-energy nucleus-nucleus collisions.

\section*{Acknowledgments}

This work is supported in part by the Natural Science Foundation of China (NSFC) under Grant Nos.~11375072 and 11647066. N-B. C. is supported by Nanhu Scholar Program for Young Scholars of XYNU. S. C. is supported by U.S. Department of Energy under Contract No. DE-SC0013460.



\end{document}